\documentclass[sigconf,anonymous=false]{acmart}

\usepackage{natbib}
\usepackage{amsmath,amsfonts}
\usepackage{array}
\usepackage{float}
\usepackage{subfig}
\usepackage{graphicx}
\usepackage{textcomp}
\usepackage{xcolor}
\usepackage[font=footnotesize]{caption}
\usepackage{multirow}

\AtBeginDocument{%
  \providecommand\BibTeX{{%
    \normalfont B\kern-0.5em{\scshape i\kern-0.25em b}\kern-0.8em\TeX}}}
\setcopyright{acmcopyright}

\copyrightyear{2023}
\acmDOI{}

\acmConference[SCF '23]{ACM Symposium on Computational Fabrication}{October 08--10,
  2023}{New York, NY}

\acmPrice{}
\acmISBN{}

\begin{document}

\title{Implicit collaboration with a drawing machine \\through dance movements}

\author{Itay Grinberg}
\affiliation{%
  \institution{Technion}
  \city{Haifa}
  \country{Israel}}
\email{ig294@cornell.edu}

\author{Alexandra Bremers}
\affiliation{%
  \institution{Cornell Tech}
  \city{New York, NY}
  \country{USA}}
\email{awb227@cornell.edu}

\author{Louisa Pancoast}
\affiliation{
\institution{louisanpancoast.com}
  \city{New York, NY}
  \country{USA}}

\author{Wendy Ju}
\affiliation{%
\orcid{0000-0002-3119-611X}
  \institution{Cornell Tech}
  \city{New York, NY}
  \country{USA}}

\renewcommand{\shortauthors}{Grinberg, et al.}

\begin{abstract}
In this demonstration, we exhibit the initial results of an ongoing body of exploratory work, investigating the potential for creative machines to communicate and collaborate with people through movement as a form of implicit interaction \cite{ju_design_2015}. The paper describes a Wizard-of-Oz demo, where a hidden wizard controls an AxiDraw drawing robot while a participant collaborates with it to draw a custom postcard. This demonstration aims to gather perspectives from the computational fabrication community regarding how practitioners of fabrication with machines experience interacting with a mixed-initiative collaborative machine. 
\end{abstract}

%\begin{CCSXML}
%\end{CCSXML}


\maketitle

\section{Introduction and Background}

%\subsection{Mixed-initiative, implicit interaction with creative machines}
Mixed-Initiative Interaction (MII), is a concept first coined by Horvitz \cite{horvitz_principles_1999} and is a paradigm that assumes that interactions are both initiated by technological artifacts as well as users---none of these two are consistently leading or following the interaction. Mixed-initiative interactions have been studied initially in the context of user interfaces \cite{horvitz_principles_1999}, and later also robotics \cite{jiang2015mixed}. 

Our research investigates what happens when mixed-initiative interaction gets brought to creative machines. Since the presence of initiative brings agency to a machine, such as a pen plotter, we can consider that machine to be a non-anthropomorphic \textit{robot}. This makes it worthwhile to consider interaction with mixed-initiative creative machines a topic in the field of human-robot interaction.

In creative activity, the activity itself can be understood as a conversation between the maker and material \cite{schon1983reflective}. Initiative from the machine itself, then, adds a \textit{third} party to the interaction---it becomes a shared interaction between a person, a machine, and the material as an ongoing reflection. When an interaction is collaborative, there is a need for communication between the interacting parties. Past human-robot interaction research has looked into the similarities between animation and robotics and the benefits of applying animation principles to the design of robotic interactions \cite{takayama_expressing_2011}, as well as ways to design for this type of interaction. Due to the collaborative nature between creative machines and people in a mixed-initiative setting, the person is not just an observer---we propose that the interaction becomes like a dance between two parties, acting on and with the material.

%\subsection{Our contribution}
In this demonstration, we present the initial results of an ongoing body of exploratory work, investigating the potential for creative machines to communicate and collaborate with people through movement as a form of implicit interaction \cite{ju_design_2015}. Through this demonstration, we aim to gather perspectives from the computational fabrication community regarding how practitioners of fabrication with machines experience interacting with a mixed-initiative collaborative machine. 

%%--------------%%
\section{The proposed demo} 
%\subsection{Setup}

%In the paper, in order to show our objective, we developed a demo between a human and a robot. We performed communication and collaboration by using the AXIDRAW movements.  

%%--------------%%
\begin{figure}%[H]
% \vspace{-.2in}
\centerline{\includegraphics[width=0.45\textwidth]{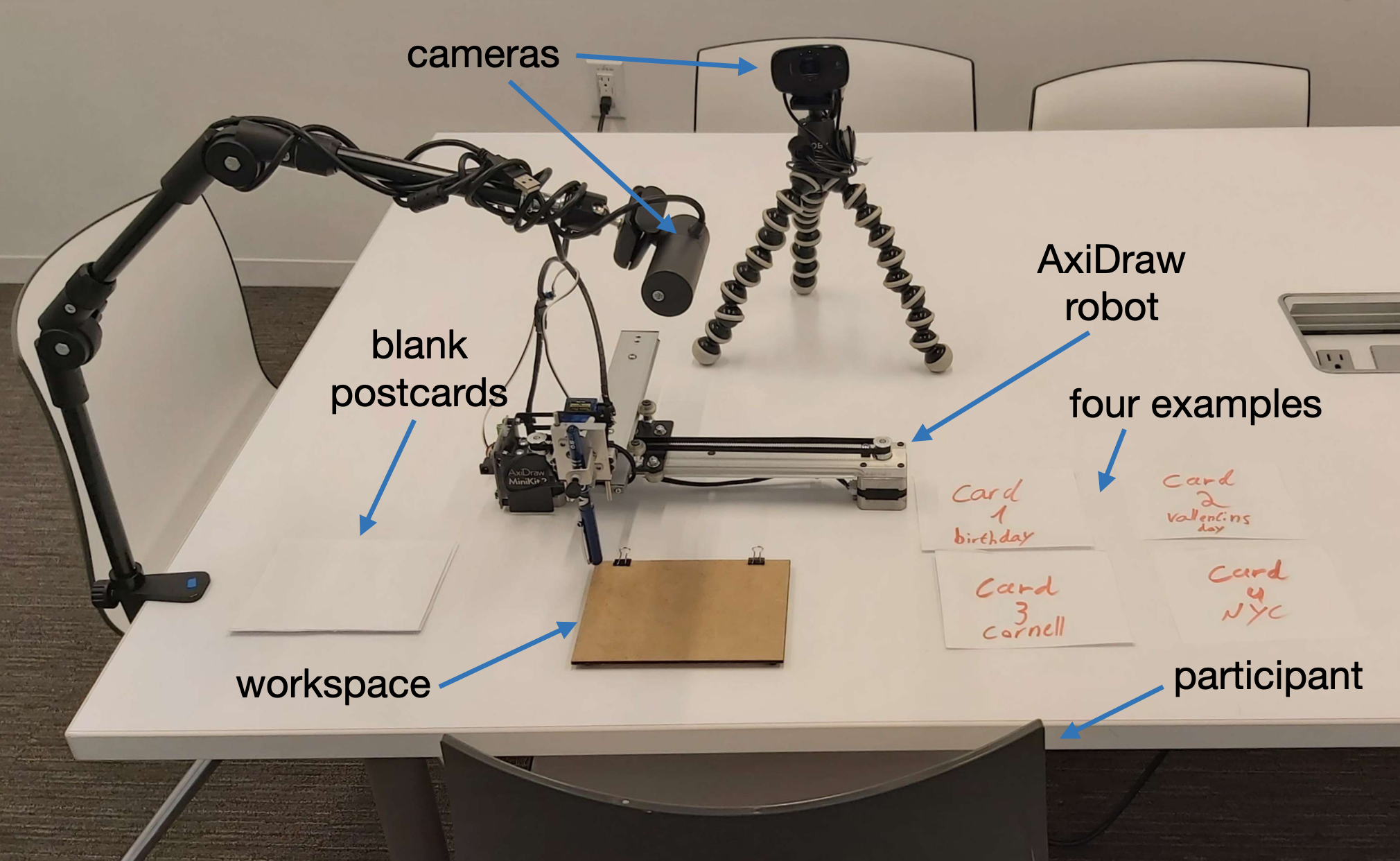}}
\caption{The demo system is built around an AxiDraw plotter, modified by adding two additional motors.}
\label{fig:setup}
% \vspace{.3in}
\end{figure}
%%--------------%%

We present a Wizard-of-Oz demo, where a hidden wizard controls a modified AxiDraw drawing robot while a participant collaborates with it to draw a custom postcard. An AxiDraw pen plotter \cite{scientist_axidraw_2022}, which is augmented with two motors adding two additional degrees of freedom, is placed on the table with two cameras aimed at the work surface and at the user. \autoref{fig:setup} shows four examples of the resulting postcards on the robot's right side. 

The Wizard will improvise using a set of pre-developed possible movements---however, the interaction consists of two clear stages that are described in \autoref{tab:welcoming}, categorized as ``Welcoming", and \autoref{tab:collaboration}, categorized as ``Collaborative Drawing". Afterwards, participants can take their postcard home.

\section{Demo Requirements}
The demo will be set up to run on a table. We require electricity and WiFi access, permission to use live cameras, and a chair for demo participants. The wizard will connect to the robot remotely by controlling the robot's computer via SSH. The wizard's ``eyes" will be two cameras -- one of them will show the participant, and the second one will show the workspace of the robot.

\section{The Demo Design}
This demonstration is intended to highlight the collaborative design method used in the design of the robot's actions. One key element of our design approach is forming our interdisciplinary research team. The authors of this demo consist of two interaction design researchers, one mechanical engineer, and one dancer. The design process occurred during in-person meetings over the course of four months. 

%%--------------%%
\begin{figure}%[H]
\centerline{\includegraphics[width=0.5\textwidth]{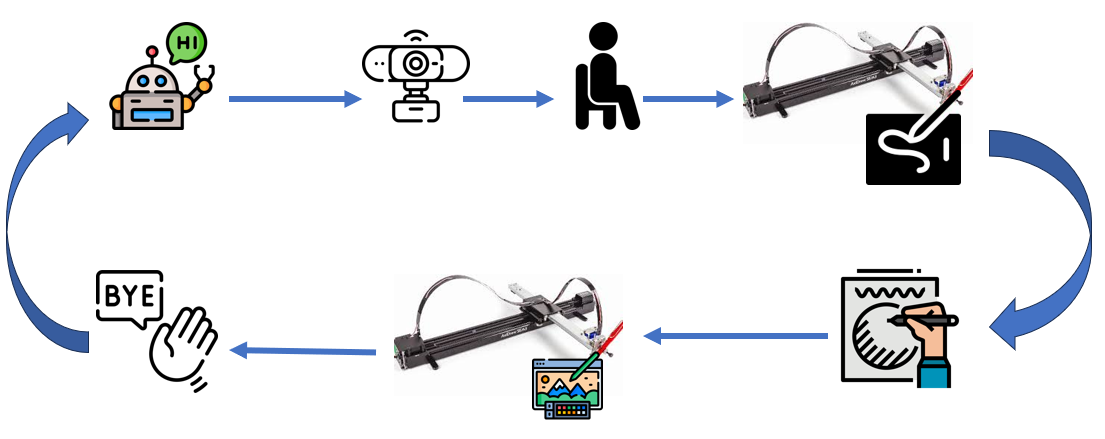}}
\vspace{-.1in}
\caption{A flow diagram of the communication scheme. \\Images \textcopyright Evil Mad Scientist Laboratories and flaticon.com (Freepik, Muhammad\_Usman, Nhor~Phai)}
\label{fig:diagram}

\end{figure}
%%--------------%%

%%--------------%%
\begin{figure}%[H]
\vspace{-.1in}
\centerline{\includegraphics[width=0.45\textwidth]{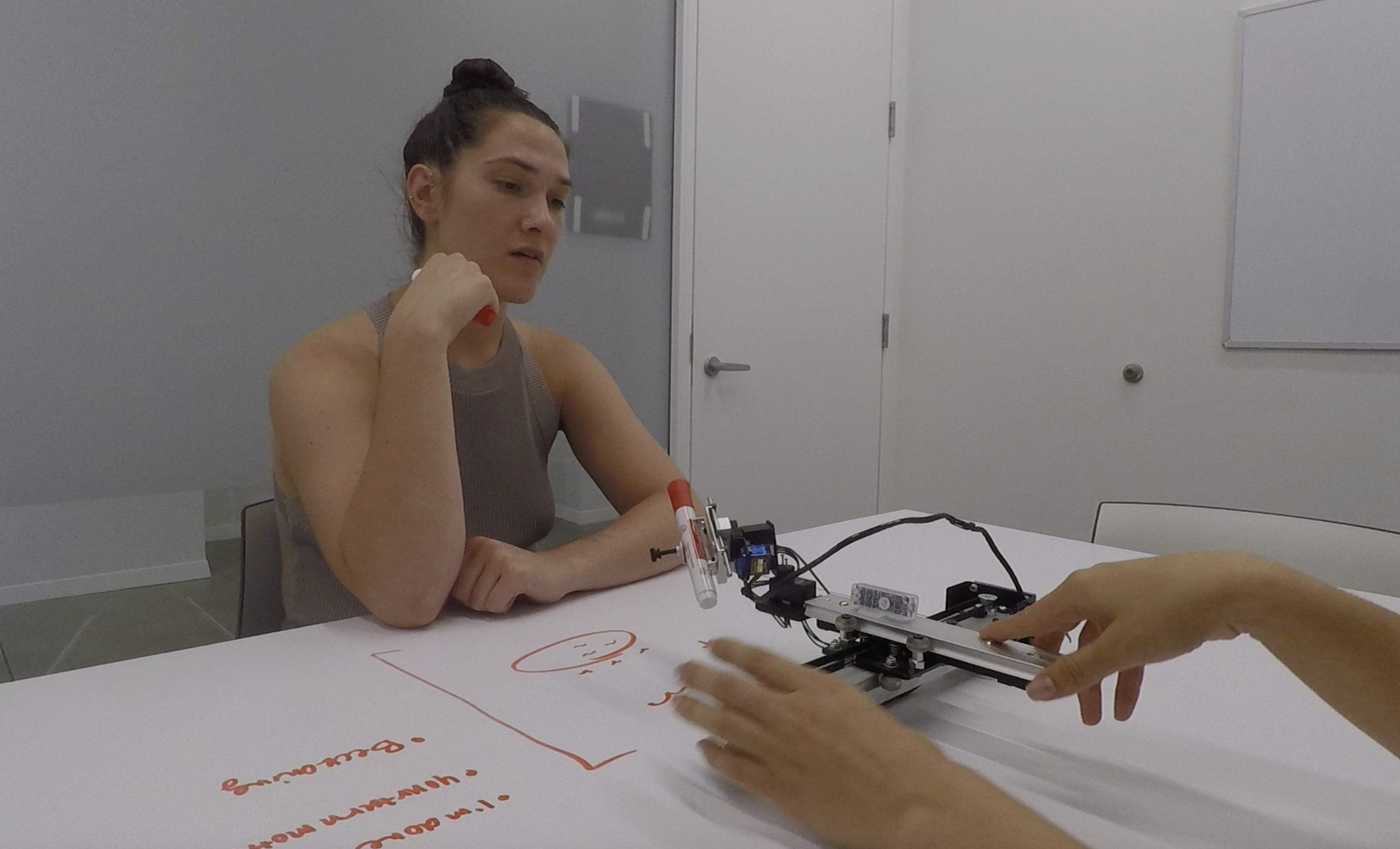}}
\vspace{-.1in}
\caption{During one design session, the dancer instructed the interaction designer how to puppeteer the movements of the AxiDraw pen plotter.}
\label{fig:process}
\vspace{-.2in}

\end{figure}
%%--------------%%

%%--------------%%
\begin{table}%[H]
    \centering
\begin{tabular}{ || m{13em} | m{12em} || } 
\hline
\multicolumn{2}{|c|}{\textbf{Stage 1: Welcoming}} \\
\hline
 \textit{Interaction (robot and you)} & \textit{Implementation (robot)} \\ 
  \hline
  The robot calls you to come here.
  The robot wants you to notice it. & Make~noise,~wave. \\ 
  \hline
  The robot makes eye contact. 
  The robot points at you. & Point pen at person and follow their gaze. \\ 
  \hline
 The robot invites you to sit down. & Point pen at person, then at chair, and back at person. \\ 
 \hline
   If the wrong person sits down, the robot tries to communicate: "No, not him -- you!"  & Shake the pen and then point at the right person. \\ 
  \hline
\end{tabular}
\caption{This table describes the potential interactions between the robot and the participant before the collaborative drawing starts.}
\label{tab:welcoming}
\end{table}

\begin{table}%[H]
\centering
\begin{tabular}{ || m{13em} | m{12.5em} || } 
    \hline
    \multicolumn{2}{|c|}{\textbf{Stage 2: Collaborative Drawing}} \\
    \hline
     \textit{Interaction (robot and you)} & \textit{Implementation (robot)} \\
    \hline  
    The robot asks what kind of card you want to make. & Point at cards and back at person. \\ 
      \hline
      The robot asks for paper. & Point at paper, participant, and back to paper. \\ 
      \hline
      The robot might start drawing. & Alternate between plotting and emotive movements. \\ 
      \hline
      The robot wants you to draw. & Point at person, at the paper, then again at the person. \\ 
      \hline
      You draw. & Wait in the home corner and make small observing movements. \\ 
      \hline
      The robot wants to see what you have drawn. & Hover over the paper before proceeding to add plotted elements. \\ 
      \hline
      The robot thinks the design is done. & Plot the "Axi" signature and then make bowing movements and wait for the participant to take the paper. \\ 
      \hline
       The robot thinks you should take the drawing away. & Point at paper, point at person, and bow again. \\ 
      \hline      
    \end{tabular}
    \caption{This table describes the potential interactions during the collaborative drawing between the robot and the participant.}
    \label{tab:collaboration}
\end{table}
%%--------------%%

%%--------------%%
\vspace{-.1in}
\section{Acknowledgements}
The authors would like to thank Cooper Murr, Tobias Weinberg, Avital Dell'Ariccia, Evil Mad Scientist, Antti Oulasvirta and François Guimbretière for their earlier suggestions, which fed into this work, and the Jacobs Technion-Cornell Institute for funding this work.

\bibliographystyle{ACM-Reference-Format}
\bibliography{bremers.bib}

\end{document}